\documentclass[conference,final,11pt]{IEEEtran}
\usepackage{url}
\usepackage{graphicx}
\usepackage{tikz}
\usepackage{amsmath} 
\usepackage{epsfig}
\usepackage{pst-grad} 
\usepackage{pst-plot} 
\renewcommand{\exp}[1]{\textrm{e}^{#1}}

\begin{document}
%
\title{ Speed-Aware Routing for UAV Ad-Hoc Networks }

 \author{
 \IEEEauthorblockN{ Stefano Rosati, Karol Kru\.zelecki, Louis Traynard, and Bixio Rimoldi.   }
 \IEEEauthorblockA{ Mobile Communications Laboratory (LCM), Information Processing Group (IPG), \\ \'Ecole Polytechnique F\'ed\'erale de Lausanne (EPFL), Lausanne,  Switzerland. 
 \\ Email: \{stefano.rosati, karol.kruzelecki, louis.traynard, bixio.rimoldi\}@epfl.ch
 }
 }
\maketitle

\IEEEpeerreviewmaketitle

\begin{abstract}
In this paper we examine mobile ad-hoc networks (MANET) composed by  unmanned aerial vehicles (UAVs). Due to the high-mobility of the nodes, these networks are very dynamic and the  existing routing protocols partly fail to provide a reliable communication.
We present Predictive-OLSR an extension to the Optimized Link-State Routing (OLSR) protocol: it enables efficient routing in very dynamic conditions. The key idea is to exploit GPS information to aid the routing protocol.  
Predictive-OLSR weights the expected transmission count (ETX) metric, taking into account the relative speed between the nodes.
We provide numerical results obtained by a MAC-layer  emulator that integrates a flight simulator to reproduce realistic flight conditions.
These numerical results show that Predictive-OLSR significantly outperforms OLSR and BABEL, providing a reliable communication even in very dynamic conditions. 
\end{abstract}

\section{Introduction} 

Unmanned aerial vehicle (UAV) networks are emerging as a valuable and suitable platform for many civilian and military applications. 
UAV networks can be used to connect users on the ground, to collect data from sensors, to provide a fast-deployable  Wi-Fi coverage in remote areas that are hardly accessible otherwise (e.g., high mountain areas).

In order to carry out challenging tasks, UAVs must be able to communicate reliably.
We show in this paper that due to the high mobility of the nodes, sometimes the  existing networks routing algorithms, which have been  designed for mobile ad-hoc networks (MANETs), such as  BABEL  \cite{bib:rfc-babel} or the Optimized Link-State Routing (OLSR) protocol \cite{bib:rfc-olsr,bib:rfc-olsr2}, fail to provide a reliable communication.

In this paper, we present an extension to OLSR, which provides reliable communication even in case of very dynamic UAV networks. 
The key idea is to exploit the GPS information. In particular, we weight the expected transmission
count (ETX) metric by a factor that takes into account the relative speed between the nodes. 

We consider an ad-hoc IEEE 802.11n network of embedded mini computers, such as the ARM-based computers produced by Gumstix Inc. \cite{bib:overoTide}, mounted on eBee drones \cite{bib:sensefly} that are produced by SenseFly. 
We test the performance of the novel algorithm by MAC-layer emulation by using the extendable mobile ad-hoc network emulator (EMANE) \cite{bib:emane} combined with eMotion 2.0 flight simulator from SenseFly.
The numerical results show that Predictive-OLSR outperforms OLSR and BABEL  and provides a reliable communication.

\section{UAV platform}

The platform is based on eBee drones \cite{bib:sensefly} developed by SenseFly and on mini ARM-based computers by Gumstix Inc. \cite{bib:overoTide}.
The drones are fixed-wing aircrafts with an electric motor and integrated autopilot capable of flying with winds of up to 12 m/s, at a cruising speed of up to 57 km/h, with an autonomy of up to 45 minutes. 
In case of emergency, they can be remotely controlled up to a distance of 3 km via a Microhard Systems Nano n2420 \cite{bib:microhard-n2420} link connection.
Within this distance, if necessary, the flight mission can be modified on the fly. The autopilot has access to an inertial measurement unit, a barometer, a pitot-tube for
airspeed, an optical-flow sensor and GPS receiver.

\begin{figure}[!htp]
	\centering
	\includegraphics[width=1.0\columnwidth]{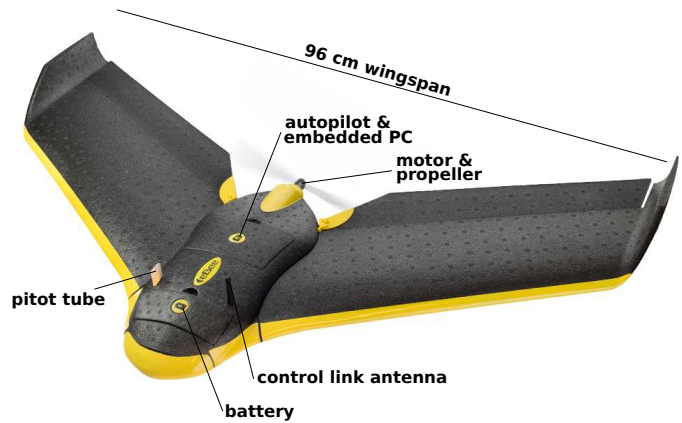} 
	\caption{SenseFly eBee drone.} 
	\label{fig:ebee}
\vspace{-0.2cm}
\end{figure}

Each eBee also carries a Gumstix Overo Tide \cite{bib:overoTide} computer with a custom  embedded Linux distribution and a standard USB WiFi (802.11n) card.
We use this embedded computer to establish the wireless network. 
A serial connection between the auto-pilot module and the embedded computer allows us to access the sensors attached to the autopilot (including the GPS reading) and to give commands to the autopilot, e.g., modify the aircraft mission according to routing needs.
Thanks to its small dimensions and weight (under 630 g), flying eBees are not considered a threat. In many countries (e.g., Switzerland) they can be used without specific authorization.

\section{Enhanced Routing for UAV Ad-Hoc Networks}

OLSR is a proactive routing algorithm based on the link-state routing protocol. It is currently the most employed routing algorithm for ad-hoc networks. 
Initially OLSR selected the route with the least number of hops \cite{bib:rfc-olsr}. 
As it is well known, the hop count metric is not suitable for wireless links.
However, using the OLSR \emph{link-quality extension} \cite{bib:olsrdLQ}, OLSR can take into account the quality of the wireless links, using the ETX metric. 

BABEL, presented more recently \cite{bib:rfc-babel}, is also a routing algorithm designed for ad-hoc networks.
Like OLSR, it is a  proactive routing algorithm, but it is based on distance-vector routing protocol and adopts EIGRP’s loop avoidance techniques.
Like OLSR, Babel also uses the ETX metric.

\subsection{Expected Transmission Count}

ETX measures the quality of the wireless link between the nodes $i$ and $j$. It was introduced in \cite{bib:Couto03ahigh-throughput} and it is defined as
\begin{equation}
 \mathrm{ETX}^{i,j}= \frac{1}{r^{\mathrm{f}}  r^{\mathrm{r}}} \, ,
\end{equation}
where $r^{\mathrm{f}}$ is the forward receiving ratio that is  the probability that a packet successfully arrives at the recipient;
and $r^{\mathrm{r}}$ is the reverse receiving ratio that is the probability that the ACK packet is successfully received.
In other words ETX estimates the expected transmissions (including re-transmissions) of a packet necessary for it to be received without error at its destination.
Then the ETX of a route $\mathcal{R}$ is defined as the sum of the ETX metrics of the links composing the route 
\begin{equation}
 \mathrm{ETX}^{\mathcal{R}}= \sum_{(i,j) \in \mathcal{R}} \mathrm{ETX}^{i,j} \, .
\end{equation}

The receiving ratios $r^{\mathrm{f}}$,  $r^{\mathrm{r}}$ are measured using as a link probe a packet called \emph{Hello packet}\footnote{In OLSR, The Hello packets are dedicated OLSR control messages that are also used as link probes by the OLSR link-quality extension}.
The \emph{Hello Interval} is a parameter that indicates how frequently the Hello packets are broadcast. 
In OLSR, the receiving ratio $r^{\mathrm{f}}$ is computed using a exponential moving average, as 

\begin{equation}
 \begin{cases}
 r^{\mathrm{f}}_{l} =  \alpha h_{l} + (1-\alpha) r^{\mathrm{f}}_{l-1} \\ 
 r^{\mathrm{f}}_{0}=0
\end{cases}
\! \! \! ,\quad 0\leq\alpha\leq1 \, ,
 \end{equation}
 where
 \begin{equation}
 h_{l} = \begin{cases}
1 & \mbox{if $l$-th Hello packet received} \\
0 & \mbox{Otherwise}  
\end{cases}
\end{equation}
and $\alpha$ is a OLSR parameter named \emph{Link-Quality Aging}.
It is worthwhile to note that $\alpha$ sets the trade-off between accuracy and responsiveness of link-quality estimation.
On one hand,  with a higher $\alpha$ the receiving ratios will be averaged for yielding  a more stable and reliable estimation;
on the other hand,   with a lower  $\alpha$ the receiving ratios will be faster for tracking the  current link-quality. 

\subsection{Speed-Weighted ETX}

ETX is an efficient measure of the quality of a link in quasi-static wireless ad-hoc networks,
but is not reactive enough  to cope with very dynamic wireless ad-hoc networks, such as UAV networks.
Due to the exponential moving average, which is  necessary for a stable and reliable link-quality estimation, 
a node takes a certain amount of time before noticing that a wireless link quality has decreased.
During this time it will continue to route packets on a broken link, thus yielding an interruption of the service.
The key idea for overcoming this problem is to use GPS information to improve the routing. 
In particular,  to predict how the link quality is likely to evolve, we use the relative speed between two nodes.
Assuming that every node knows the position of its neighbors\footnote{In the following section we discuss how to distribute this information among the nodes.},
we redefine ETX as

\begin{equation}
 \mathrm{ETX}^{i,j} = \frac{\exp{v_{\ell}^{i,j}  \beta}}{r^{\mathrm{f}}  r^{\mathrm{r}}} \, ,
\end{equation}
where $v_{\ell}^{i,j}$ is the relative speed between node $i$ and $j$, and  $\beta$ is a non-negative parameter.

If the node $i$ and $j$ move closer to each other, the relative speed is negative, thus the ETX will be weighted by a factor smaller than 1. 
On the contrary, if the node $i$ and $j$ move away from each other, the relative speed is positive, thus the ETX will be weighted by a factor greater than 1. 
As a consequence, a link between two nodes that move closer will be preferred to a link between two nodes that move apart, even if they have the same values of $r^{\mathrm{f}}$ and $r^{\mathrm{r}}$.
The best value of $\beta$ depends on the  cruise speed of the UAVs of the link coverage extension.

\subsection{Computation of the Speed}
As discussed in the following section, in our implementation the GPS coordinates are conveyed by the Hello packets. 
Thus every time the algorithm computes the ETX, it has available fresh GPS information.
The instantaneous relative velocity between $i$ and $j$ at time $t_i$ is computed as
\begin{equation}
 \tilde{v}_{\ell}^{i,j}= \frac{ d_{\ell}^{i,j} - d_{\ell-1}^{i,j} }{t_{\ell} - t_{\ell-1}} \, ,
\end{equation} 
where, $t_{\ell}$ and $t_{\ell-1}$ are, respectively, the arrival time of the last and second to last Hello packet received.  $d_{\ell}^{i,j}$ $d_{\ell-1}^{i,j}$ are the corresponding distances between the nodes $i$ and $j$. 
As the GPS positions are subject to errors, and gusts of wind can perpetuate the  motion of the small UAV, it is preferable to average the instantaneous speed using  a exponential moving average as follows
\begin{equation}
 v_{\ell}^{i,j} =  \gamma \tilde{v}_{\ell}^{i,j}  + (1-\gamma) v_{\ell-1}^{i,j} ,\quad 0\leq\gamma\leq1 \, ,
 \end{equation}
where $\gamma$ is a Predictive-OLSR parameter.

\section{Implementation Details}

To implement this speed-weighed ETX, we used the already-existing, open-source and actively developed OLSR daemon called OLSRd \cite{bib:olsrd}. We modified it to share position information in addition to receiving ratios in Hello packets. Thus, every node knows its neighbor's position and can compute the corresponding ETX.

\subsection{OLSRd Link-Quality Extension}

OLSRd uses link-quality sensing and ETX metrics, through the so-called Link Quality extension \cite{bib:olsrdLQ}. They replace the hysteresis mechanism of the OLSR protocol with link-quality sensing algorithms, intended to be used with ETX-based metrics. To do so, such an extension uses OLSR Hello messages to probe links quality and advertise link-specific quality information (receiving ratios), in addition to their previous role in the protocol (detecting and advertising neighbors). Likewise, the link-quality extension include link-quality information in OLSR TC messages, to distribute it to the whole network.

Clearly the modified messages are not RFC-compliant anymore, because they include new fields for link-quality information. 
Therefore, all the nodes in the network should use the link-quality extension.

\subsection{Extending Link-Quality Extension}

Position information can be thought as additional link-quality information, which is shared the same way as receiving ratios.
The OLSRd link-quality extension mechanism was tailored for simple ETX metrics.
Now, in our case, we have two new requirements:  (i) store more information per link to compute the ETX; (ii) to share the position information that is not link-specific.
We extended the OLSR link-quality extension mechanism to enable the implementation of more complex metrics.
In particular, we modified again the Hello message to include the GPS positions that are non-link-specific information. 
The modified format of Hello packets is reported in Figure \ref{fig:hello}. 
The fields in gray are not part of the original OLSR RFC.
The fields \emph{neighbor-specific link-quality data} have been introduced by the OLSRd link-quality extension. 
We added a new field named \emph{non-link-specific quality data}.

\subsection{Obtaining GPS Position}
OLSRd comes with a handy networking toolkit that we used to implement an OLSRd plug-in able to listen to GPS sentences on a given interface, parse them (with the open-source NMEA library) and update the node position, directly in OLSRd.

\begin{figure}
\centering
\scalebox{0.8} 
{
\begin{pspicture}(0,-5.34)(10.0,5.34)
\definecolor{color57b}{rgb}{0.7529411764705882,0.7450980392156863,0.7450980392156863}
\psframe[linewidth=0.04,dimen=outer](10.0,5.5)(0.0,4.54)
\psframe[linewidth=0.04,dimen=outer](10.0,4.58)(0.0,3.78)
\psframe[linewidth=0.04,dimen=outer,fillstyle=solid,fillcolor=color57b](10.0,3.82)(0.0,3.02)
\psframe[linewidth=0.04,dimen=outer](10.0,3.06)(0.0,2.26)
\psframe[linewidth=0.04,dimen=outer](10.0,2.32)(0.0,1.52)
\psframe[linewidth=0.04,dimen=outer,fillstyle=solid,fillcolor=color57b](10.0,1.56)(0.0,0.76)
\psframe[linewidth=0.04,dimen=outer](10.0,0.8)(0.0,0.0)
\psframe[linewidth=0.04,dimen=outer,fillstyle=solid,fillcolor=color57b](10.0,0.04)(0.0,-0.76)
\usefont{T1}{ppl}{m}{n}
\rput(5.12625,1.1433333){Neighbor-specific link-quality data}
\usefont{T1}{ppl}{m}{n}
\rput(5.12625,0.39666668){Neighbor interface address}
\usefont{T1}{ppl}{m}{n}
\rput(5.12625,-0.35){Neighbor-specific link-quality data}
\usefont{T1}{ppl}{m}{n}
\rput(5.1303124,-1.155){\Large ...}
\psframe[linewidth=0.04,dimen=outer](10.0,-0.72)(0.0,-1.52)
\usefont{T1}{ppl}{m}{n}
\rput(5.12625,1.89){Neighbor interface address}
\usefont{T1}{ppl}{m}{n}
\rput(1.1934375,2.635){Link Code}
\usefont{T1}{ppl}{m}{n}
\rput(7.6034374,2.67){Link Message Size}
\usefont{T1}{ppl}{m}{n}
\rput(3.6903124,2.635){Reserved}
\psline[linewidth=0.04cm](2.4,3.02)(2.4,2.3)
\psline[linewidth=0.04cm](5.0,3.06)(5.0,2.3)
\usefont{T1}{ppl}{m}{n}
\rput(1.57,4.8){\large 0 1 2 3 4 5 6 7 8 9}
\usefont{T1}{ppl}{m}{n}
\rput(8.066406,4.8){\large0 1 2 3 4 5 6 7 8 9 0 1}
\usefont{T1}{ppl}{m}{n}
\rput(4.67,4.8){\large0 1 2 3 4 5 6 7 8 9}
\usefont{T1}{ppl}{m}{n}
\rput(0.160625,5.2){\large 0}
\usefont{T1}{ppl}{m}{n}
\rput(6.2940625,5.2){\large 2}
\usefont{T1}{ppl}{m}{n}
\rput(3.269375,5.2){\large 1}
\usefont{T1}{ppl}{m}{n}
\rput(9.476875,5.2){\large 3}
\psline[linewidth=0.04cm](4.98,4.56)(4.98,3.8)
\psline[linewidth=0.04cm](7.5,4.56)(7.5,3.82)
\usefont{T1}{ppl}{m}{n}
\rput(2.3703125,4.155){Reserved}
\usefont{T1}{ppl}{m}{n}
\rput(6.1534376,4.155){Htime}
\usefont{T1}{ppl}{m}{n}
\rput(8.769062,4.19){Willingness}
\usefont{T1}{ppl}{m}{n}
\rput(5.14625,3.43){Non-link-specific quality data}
\psframe[linewidth=0.04,dimen=outer](10.0,-1.5)(0.0,-2.3)
\psframe[linewidth=0.04,dimen=outer,fillstyle=solid,fillcolor=color57b](10.0,-2.26)(0.0,-3.06)
\psframe[linewidth=0.04,dimen=outer](10.0,-3.02)(0.0,-3.82)
\psframe[linewidth=0.04,dimen=outer,fillstyle=solid,fillcolor=color57b](10.0,-3.78)(0.0,-4.58)
\usefont{T1}{ppl}{m}{n}
\rput(5.12625,-2.6766667){Neighbor-specific link-quality data}
\usefont{T1}{ppl}{m}{n}
\rput(5.12625,-3.4233334){Neighbor interface address}
\usefont{T1}{ppl}{m}{n}
\rput(5.12625,-4.17){Neighbor-specific link-quality data}
\usefont{T1}{ppl}{m}{n}
\rput(5.1303124,-4.975){\Large ...}
\psframe[linewidth=0.04,dimen=outer](10.0,-4.54)(0.0,-5.34)
\usefont{T1}{ppl}{m}{n}
\rput(5.12625,-1.93){Neighbor interface address}
\end{pspicture} 
}
\caption{Format of the modified Hello packet.}
\label{fig:hello}
\vspace{-0.2cm}
\end{figure}

\section{Routing Performance}
We measure the evolution of the datagram loss rate (DLR) of a multi-hop route.
As we use the network to transmit a continuous stream of data (e.g., high-quality video stream) in real time, our goal is to minimize the DLR during the transmission.
We measure the DLR every second by sending 85 UDP datagrams having, in total, a size of 1 Mbit\footnote{We use \emph{iperf} to obtain this measurements.}. 
DLR is the ratio between the lost and the total number of  datagrams.
We consider two different multiple-hop scenarios: (i) involving a UAV source node, two UAV relays, and a fixed destination node; (ii) involving a UAV node flying around rectangle area sized 1200x1500 meters  and covered by 32 static relays.
We compare Predictive-OLSR, OLSR that uses the link-quality extension, and BABEL.
We set the Hello interval to 0.5 second for all the analyzed algorithms.
For OLSR we set $\alpha=0.2$, which is the best trade-off between stability and responsiveness in our scenarios.
As Predictive-OLSR is inherently more responsive, we can adopt a lower $\alpha$ in order to improve the stability. We choose $\alpha=0.05$.
The others Predictive-OLSR parameters are set as follows: $\beta=0.2$, $\gamma=0.04$.
All the  measurements are obtained using the MAC-layer emulator presented  in the following section.

\subsection{Emulation Platform}

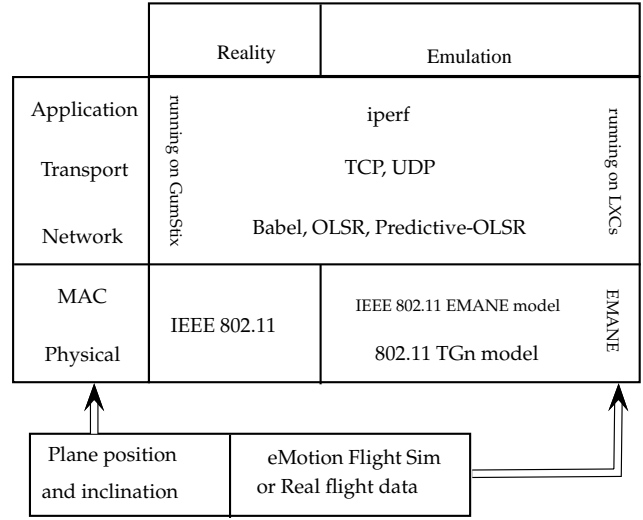
\begin{figure}
\scalebox{0.7} 
{
\begin{pspicture}(0,-4.91)(11.98,4.91)
\usefont{T1}{ppl}{m}{n}
\rput(1.3723438,2.84){Application}
\usefont{T1}{ppl}{m}{n}
\rput(1.3760937,1.6804688){Transport}
\usefont{T1}{ppl}{m}{n}
\rput(1.3398438,0.475){Network}
\usefont{T1}{ppl}{m}{n}
\rput(7.1629686,0.6175){Babel, OLSR, Predictive-OLSR}
\usefont{T1}{ppl}{m}{n}
\rput(7.184219,1.7525){TCP, UDP}
\usefont{T1}{ppl}{m}{n}
\rput(7.1492186,2.734375){iperf}
\usefont{T1}{ppl}{m}{n}
\rput(1.3398438,-0.686875){MAC}
\usefont{T1}{ppl}{m}{n}
\rput(1.3398438,-1.82){Physical}
\usefont{T1}{ppl}{m}{n}
\rput{-90.0}(9.746875,13.063437){\rput(11.376094,1.665){\footnotesize running on LXCs}}
\usefont{T1}{ppl}{m}{n}
\rput(8.46875,-0.846875){\footnotesize IEEE 802.11 EMANE model}
\usefont{T1}{ppl}{m}{n}
\rput(8.39875,-1.755){ 802.11 TGn model}
\psframe[linewidth=0.04,dimen=outer](11.94,4.91)(2.58,3.49)
\psframe[linewidth=0.04,dimen=outer](2.62,3.51)(0.0,-2.33)
\psframe[linewidth=0.04,dimen=outer](11.94,3.51)(2.58,-2.33)
\psframe[linewidth=0.04,dimen=outer](8.72,-3.27)(0.32,-4.87)
\psline[linewidth=0.04cm](0.04,-0.07)(11.94,-0.07)
\psline[linewidth=0.04cm](5.86,-0.07)(5.86,-2.35)
\psline[linewidth=0.04cm](5.86,4.87)(5.86,3.53)
\psline[linewidth=0.04cm](4.14,-3.31)(4.14,-4.89)
\usefont{T1}{ppl}{m}{n}
\rput(1.9073437,-3.72){Plane position}
\usefont{T1}{ppl}{m}{n}
\rput(6.50625,-3.78){eMotion Flight Sim}
\usefont{T1}{ppl}{m}{n}
\rput(1.8579688,-4.38){ and inclination }
\usefont{T1}{ppl}{m}{n}
\rput(6.180156,-4.34){or Real flight data}
\usefont{T1}{ppl}{m}{n}
\rput{-90.0}(12.6,10.253437){\rput(11.395156,-1.195){\footnotesize EMANE}}
\psline[linewidth=0.021999998cm,arrowsize=0.052916665cm 2.0,arrowlength=1.4,arrowinset=0.4,doubleline=true,doublesep=0.12,doublecolor=white]{->}(1.56,-3.29)(1.56,-2.33)
\psline[linewidth=0.021999998,arrowsize=0.052916665cm 2.0,arrowlength=1.4,arrowinset=0.4,doubleline=true,doublesep=0.12,doublecolor=white]{->}(8.74,-4.05)(11.5,-4.03)(11.52,-2.33)
\usefont{T1}{ptm}{m}{n}
\rput(4.4534373,3.92){Reality}
\usefont{T1}{ptm}{m}{n}
\rput(8.691093,3.88){Emulation}
\usefont{T1}{ppl}{m}{n}
\rput(4.0153127,-1.231875){IEEE 802.11}
\usefont{T1}{ppl}{m}{n}
\rput{-90.0}(1.3907812,4.747344){\rput(3.04,1.685){\footnotesize running on GumStix}}
\end{pspicture} 
}
\caption{Emulation platform.}
\label{fig:emulator}
\vspace{-0.2cm}
\end{figure}

Field experiments are expensive and require the involvement of people, transportation, and costly equipment. For this reason we developed an emulation platform that integrates all the test-bed aspects, as illustrated in Figure \ref{fig:emulator}. 
The emulator creates a  Linux Container (LXC) for each node of the network.  The nodes are connected using a MAC-layer emulator called extendable mobile ad-hoc network emulator (EMANE). EMANE is an open-source framework, developed mainly by Naval Research Laboratory for real-time modeling of mobile network systems.  
Regarding the channel model,  we consider the IEEE 802.11 TGn model defined in  \cite{bib:802.11TGn}. 
EMANE imports the positions, speeds, and orientations of the UAVs from log files. 
These log files can be obtained from real flight data logs, or  by a flight simulator called eMotion, provided by SenseFly, that simulates realistic flight condition of the eBee drone. 
All network layers, except the MAC and the physical layer, use the actual implementations that run in the Linux machine hosting the emulation.
To obtain numerical results we run emulation on Fedora 15, kernel 2.6.43.8. The tested version of OLSRd  is 0.6.5.3, and the tested version of BABELd is 1.3.4.

\subsection{2-relay scenario}

\begin{figure}[!htp]
	\centering
	\includegraphics[width=1.0\columnwidth]{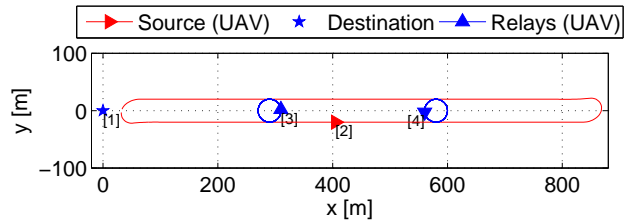} 
	\vspace{-0.2cm}
	\caption{2-relay scenario.}
	\label{fig:2relays}
	\vspace{-0.2cm}
\end{figure}

A multiple-hop scenario, illustrated in Figure \ref{fig:2relays}, consists of a mobile UAV source (node 2), two mobile UAV relays (nodes 3 and 4), and a fixed destination (node 1). Both the source and the relay nodes are embedded in eBee UAVs.
The relays keep circling around the given position with the circular trajectories of 20 meter radius, extending the range of the network coverage for the mobile source from around 300 meters up
to around 900 meters from the target. 
Node 2 follows a straight trajectory of around 850 meters, then it returns to the starting point.  It takes 160 seconds to finish the loop. During this period, node 2 continuously transmits  data to node 1.  
During one loop, node 2 changes its routing to node 1  from a direct connection to a 2-hop and then to a 3-hop route, and back. Therefore the network topology will change 4 times during the loop.

\begin{figure}
\includegraphics[width=1.0\columnwidth]{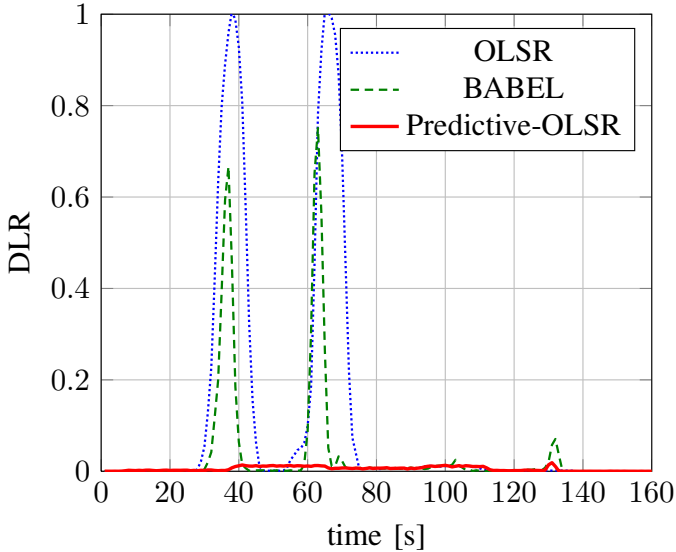} 
 \caption{Evolution of average DLR in the 2-relay scenario.}
 \label{fig:2relays-per}
 \vspace{-0.2cm}
\end{figure}

For each routing algorithm, we run 200 loops in order to obtain an average of the results.  
Figure  \ref{fig:2relays-per} shows the evolution of the average  DLR during the loop. 
For both OLSR and BABEL, we  notice  the two peaks of the DLR that correspond  to the moments when the routing algorithm has to switch from the direct link to a 2-hop and then to a 3-hop connection.
This happens because  the routing algorithm takes a certain amount of time to notice that the wireless direct-link quality is broken. So it reacts late, this translates into an interruption of the service.
Whereas, Predictive-OLSR reacts promptly to the topology changes. Figure  \ref{fig:2relays-per} shows  that there are no peaks of the average DLR.
It is interesting to note that BABEL outperforms OLSR. Similar results were reported \cite{bib:OlsrBABELComparison} where BABEL and OLSR were experimentally compared.

\begin{figure}
	\centering
	\includegraphics[width=1.0\columnwidth]{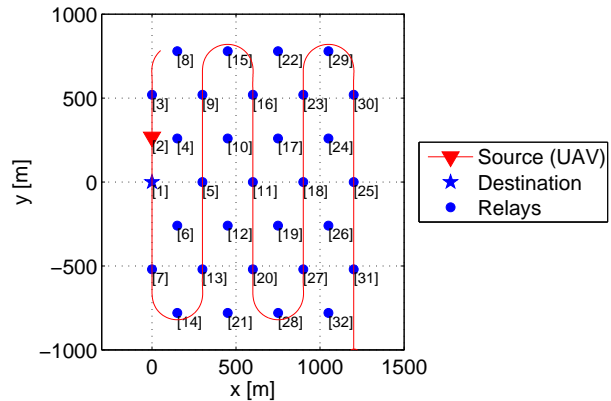} 
	\vspace{-1.2cm}
	\caption{Open area coverage scenario.}
	\label{fig:32relays}
	\vspace{-0.2cm}
\end{figure}

\begin{figure*}
\centering
\includegraphics[width=2.0\columnwidth]{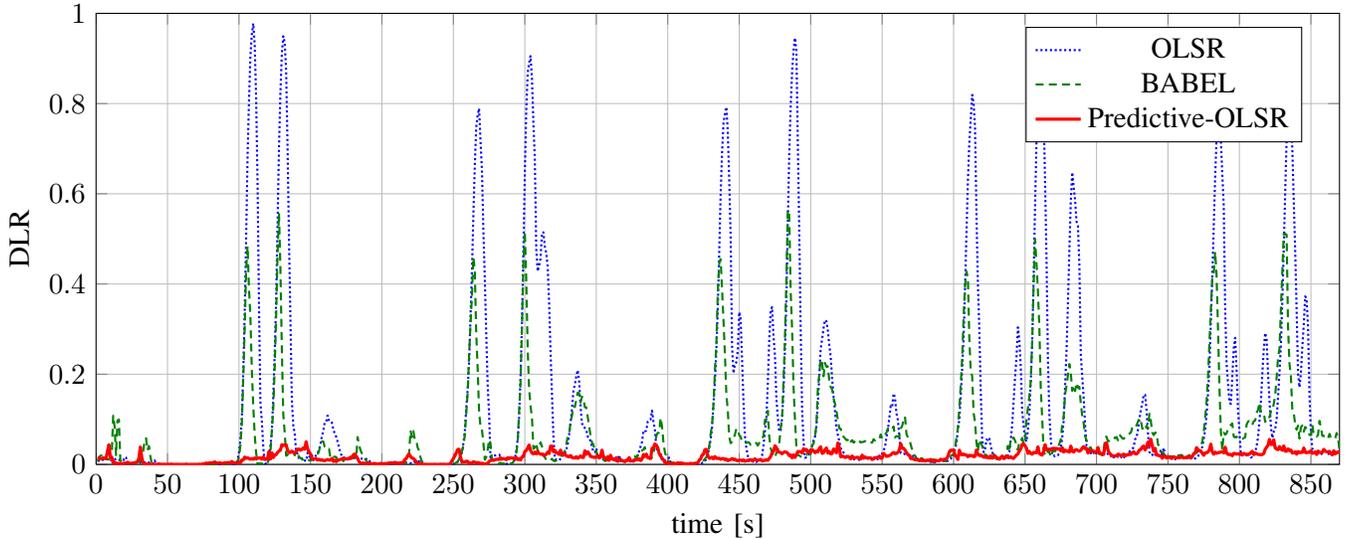} 
 \caption{Evolution of average DLR in the open area coverage scenario.}
 \label{fig:32relays-per}
 \vspace{-0.2cm}
\end{figure*}

\begin{figure}
\includegraphics[width=1.0\columnwidth]{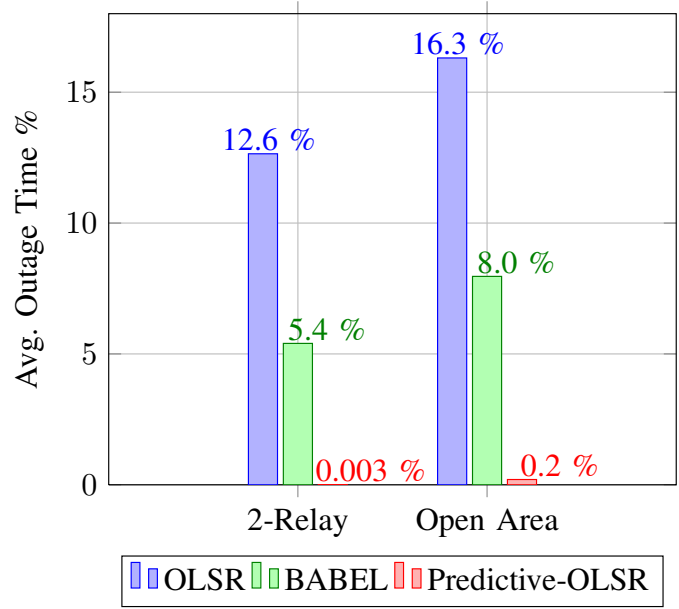} 
\caption{Average outage time for both the analyzed scenarios.}
\label{fig:outage}
\vspace{-0.2cm}
\end{figure}

\subsection{Open Area Coverage}

In the second scenario, consisting of one moving UAV (node 2), one fixed destination (node 1) and 30 relays (nodes 3 to 32) that are located uniformly within a rectangle area  of size 1200x1500 meters.
As before, we transmit the data continuously from node 2 to node 1. Node 2 scans the area by following the trajectory shown in Figure \ref{fig:32relays}. It takes 870 seconds to complete the trajectory. 
The distances between the relays has been chosen to have a good quality direct wireless link only among the closest neighbors that are 300 meters away. For example, node 5 can communicate directly only with nodes 1, 4, 6, 10, 11, 12. It cannot reach the other nodes directly.  
In order to average the results, we repeat the emulation 50 times for each routing algorithm.

Figure  \ref{fig:32relays-per} shows the evolution of the average  DLR. 
Again, we notice that, for  both  OLSR and BABEL, the DLR has several peaks during the mission, which translates into a  service interruption.  This happens because the routing algorithm is not fast enough to reach to the topology changes. 
Whereas, with Predictive-OLSR the average DLR is never higher than 0.1. 

Let us assume that there is an outage event during the time the DLR is greater than 0.2. We compute the outage time  percentage for each run, and then we average all the runs. As shown in Figure \ref{fig:outage}, node 2 experiences an average outage time of 16.3\% with OLSR: and 8\% with BABEL; where as with Predictive-OLSR the outage time is  0.2\%.

\section{Conclusions}

In this paper, we have presented an extension, named Predictive-OLSR, to the OLSR routing protocol, that enables efficient routing in very dynamic ad-hoc networks composed of UAVs. 
This extension exploits GPS information. 
The numerical results, obtained by MAC-layer emulation, show that Predictive-OLSR succeeds in providing a reliable multi-hop communication, even in such a dynamic ad-hoc network, whereas, other state-of-the-art  routing protocols, such as BABEL and OLSR, mostly fail.

\section*{Acknowledgments} This work is supported by armasuisse, competence sector. Science+Technology for the Swiss Federal Department of Defense, Civil Protection and Sport. 

\bibliographystyle{IEEEtran} 
\bibliography{smavnet,IEEEfull}
\addcontentsline{toc}{part}{\small References}
\end{document}